\documentclass[epj]{webofc}
\usepackage[varg]{txfonts}   
\wocname{EPJ Web of Conferences}
\woctitle{INPC 2013}
\begin{document}
\title{How neutron stars constrain the nuclear equation of state}
%
%

\author{Thomas Hell\inst{1,2}\fnsep\thanks{\email{thell@ph.tum.de}} \and
       Bernhard R\"ottgers\inst{1} \and
        Wolfram Weise\inst{1,2}
}

\institute{Physik Department, Technische Universit\"{a}t M\"{u}nchen, D-85747 Garching, Germany
\and
	ECT*, Villa Tambosi, I-38123 Villazzano (Trento), Italy
          }

\abstract{Recent neutron star observations set new constraints for the equation of state of baryonic matter. A chiral effective field theory approach is used for the description of neutron-dominated nuclear matter present in the outer core of neutron stars. Possible hybrid stars with quark matter in the inner core are discussed using a three-flavor Nambu--Jona-Lasinio model.
}
\maketitle
\section{Introduction}
\label{intro}
The masses of the pulsars J1614-2230 and J0348+0432 have recently been determined with high
accuracy as $1.97\pm0.04$ \cite{Demorest:2010bx} and $2.01\pm0.04$ \cite{Antoniadis:2013pzd} solar masses, respectively. Combined with (less accurate) statistical analyses of neutron star radii \cite{Lattimer:2010uk,Steiner:2010fz,Steiner:2012xt,Lattimer:2013hma,Guillot:2013wu,Truemper2013} giving an allowed range of about $10$~--~$15\,{\rm km}$, this leads to tight constraints for the equation of state (EoS) of dense baryonic
matter inside neutron stars. Masses and radii of static and isotropic neutron stars are determined from the Tolman-Oppenheimer-Volkoff equations. Equations of state based on chiral effective field theory (ChEFT) and on a Nambu--Jona-Lasinio (NJL) model are described in Sec.~\ref{sec-3} and combined in Sec.~\ref{sec-4} with a realistic phenomenological equation of state for the neutron star crust at low densities in order to discuss several scenarios for neutron star matter. For the NJL quark-matter EoS we investigate the role of a vector current interaction for the chiral phase transition.  
Results are summarized and discussed in Sec.~\ref{sec-5}.

\section{Modeling the equation of state of neutron stars}\label{sec-3}

Two complementary approaches to the equation of state (EoS) for neutron star matter are used in this work: chiral effective field theory (ChEFT), and the Nambu--Jona-Lasinio (NJL) model. The resulting EoS's imply different scenarios for the composition of neutron star core regions (see Sec.~\ref{sec-4}). Generic constraints on the EoS in view of recent observations include (cf.~\cite{Hebeler:2010jx,Hebeler:2013nza}) a phenomenological crust EoS \cite{Haensel:2004nu,Douchin:2001sv} up to $\varrho\lesssim0.8\,\varrho_0$ ($\varrho_0=0.16\,{\rm fm}^{-3}$), the nuclear-matter EoS derived in Ref.~\cite{Fiorilla:2011sr} in the nuclear terrain around $\varrho_0$, and an extrapolation to higher densities using three polytropes, $P=K_i\,\varrho^{\varGamma_i}$, $i\in\{1,2,3\}$. The exponents in the polytropes are constrained by a) the requirement that the resulting EoS must support two-solar-mass neutron stars; b) the causality condition; c) the resulting neutron star radii should be in the acceptance range reported in either Refs.~\cite{Steiner:2010fz,Steiner:2012xt} (green band in Figs.~\ref{fig-eoschpt} and \ref{fig-njlgibbs}) or in Ref.~\cite{Truemper2013} (blue band).

{\bf Chiral effective field theory}\label{sec-31}. The EoS of nuclear and neutron star matter based on ChEFT is derived  following Ref.~\cite{Kaiser:2001jx}.
Starting from the next-to-leading-order (NLO) chiral meson-baryon effective Lagrangian, the resulting EoS incorporates all one- and two-pion exchange processes in the medium up to three-loop order in the free-energy density (i.e., to order $\mathcal{O}(k_F/M_N)^5$ where $k_F$ denotes the nuclear Fermi momentum und $M_N$ is the nucleon mass). Contact terms subsuming all unresolved short-distance nucleon-nucleon dynamics and three-nucleon interactions are properly considered. A  resummation of in-medium ladder diagrams involving nucleon-nucleon contact interactions to all orders in the scattering length $a$ is performed \cite{Kaiser:2011cg}. This framework is used to determine the energies per particle for both symmetric nuclear matter (SNM) and pure neutron matter (PNM).


{\bf NJL model with vector interaction}\label{sec-32}. The possibility of hybrid stars with a quark matter core is discussed in terms of an EoS constructed from a three-flavor Nambu--Jona-Lasinio (NJL) model with an additional repulsive vector interaction term with strength $G_v$. This NJL Lagrangian, $\mathcal{L}=\bar q(x)\left(-{\rm i}\,\gamma^\mu\partial_\mu+\gamma^0\hat\mu+\hat m\right)q(x)+\mathcal{L}_{\rm int}\,,$ has an interaction part \cite{Vogl:1991qt,Klimt:1989pm}:
\begin{equation}\label{3flint}
	\mathcal{L}_{\rm int}=\dfrac{1}{2}\, G\sum_{a=0}^8\left[(\bar q\lambda^a q)^2+(\bar q{\rm i}\gamma_5\lambda^a q)^2\right]-K\left[\det\left(\bar q(1+\gamma_5) q\right)+\det\left(\bar q(1-\gamma_5) q\right)\right]-\dfrac{1}{2}\,G_v\,(\bar q\gamma^\mu q)^2\,.
	\end{equation}
Here, $q(x)=(u(x),d(x),s(x))^\top$ is the three-flavor quark field, $\hat m={\rm diag}_f(m_u,m_d,m_s)$ denotes the current-quark mass matrix (setting $m_u=m_d$ in the isospin limit), and $\hat\mu={\rm diag}_f(\mu_u,\mu_d,\mu_s)$ is the quark chemical-potential matrix in flavor space. Apart from the standard NJL coupling $G$, the 't~Hooft-Kobayashi-Maskawa determinant interaction with strength $K$ takes into account the anomalous breaking of the axial ${\rm U}(1)$ symmetry.
Calculations are performed  in mean-field (MF) approximation.

\section{Constraints from neutron stars}\label{sec-4}

For the inhomogeneous matter forming the neutron star crust we use (at densities below about $0.8\,\varrho_0$) the phenomenological SLy EoS \cite{Haensel:2004nu,Douchin:2001sv}. Charge neutrality and beta equilibrium ($n\leftrightarrow p+e^-+\bar\nu_e$, $n\leftrightarrow p+\mu^-+\bar\nu_\mu$) imply relations between the chemical potentials or the densities appearing in the EoS described in Sec.~\ref{sec-3}. Electrons and muons are treated as free Fermi gases. 

\begin{figure}
\begin{minipage}{.475\textwidth}\vspace{.2cm}
		\includegraphics[width=\textwidth]{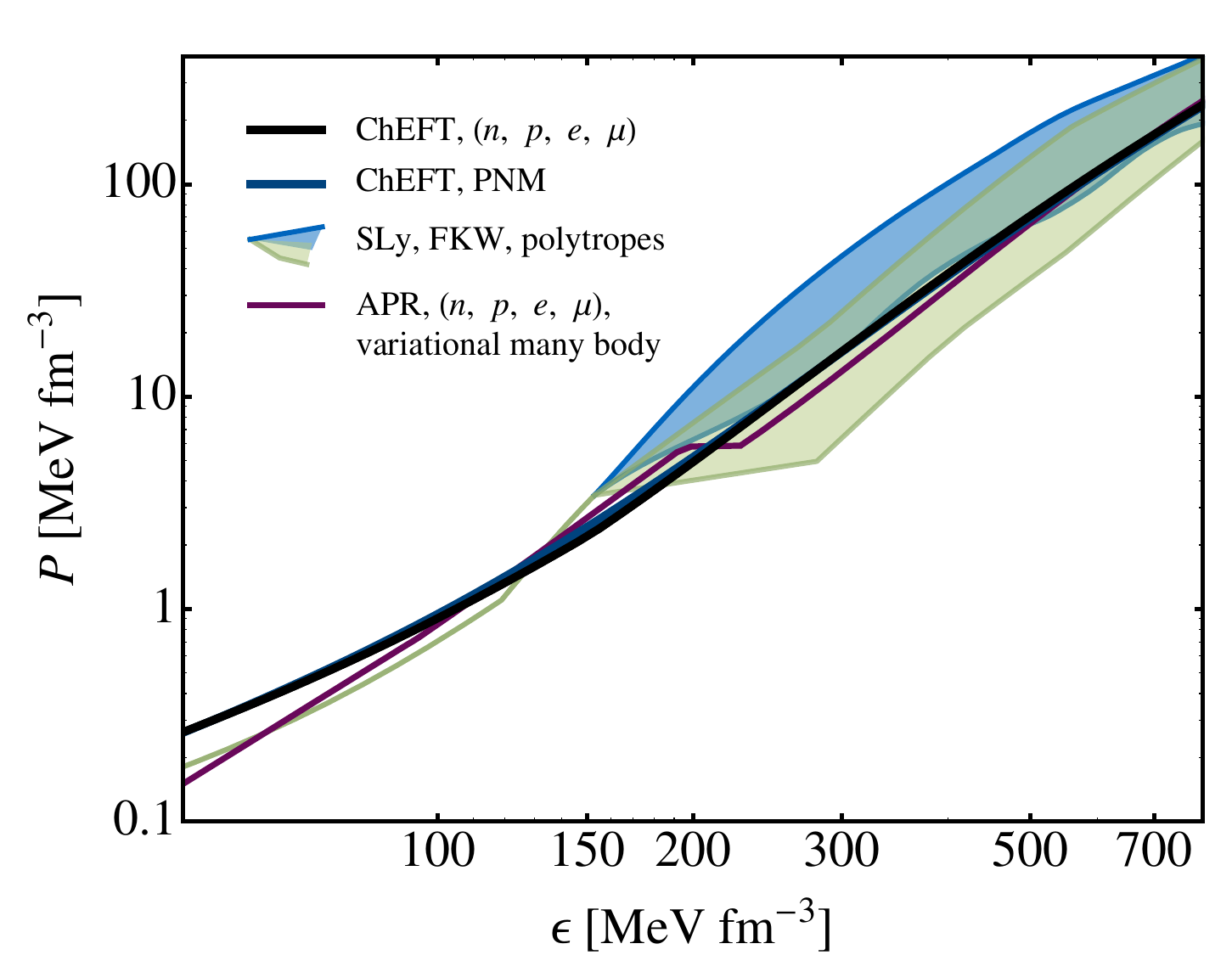}
	\end{minipage}
	\hfill
\begin{minipage}{.485\textwidth}
		\includegraphics[width=\textwidth]{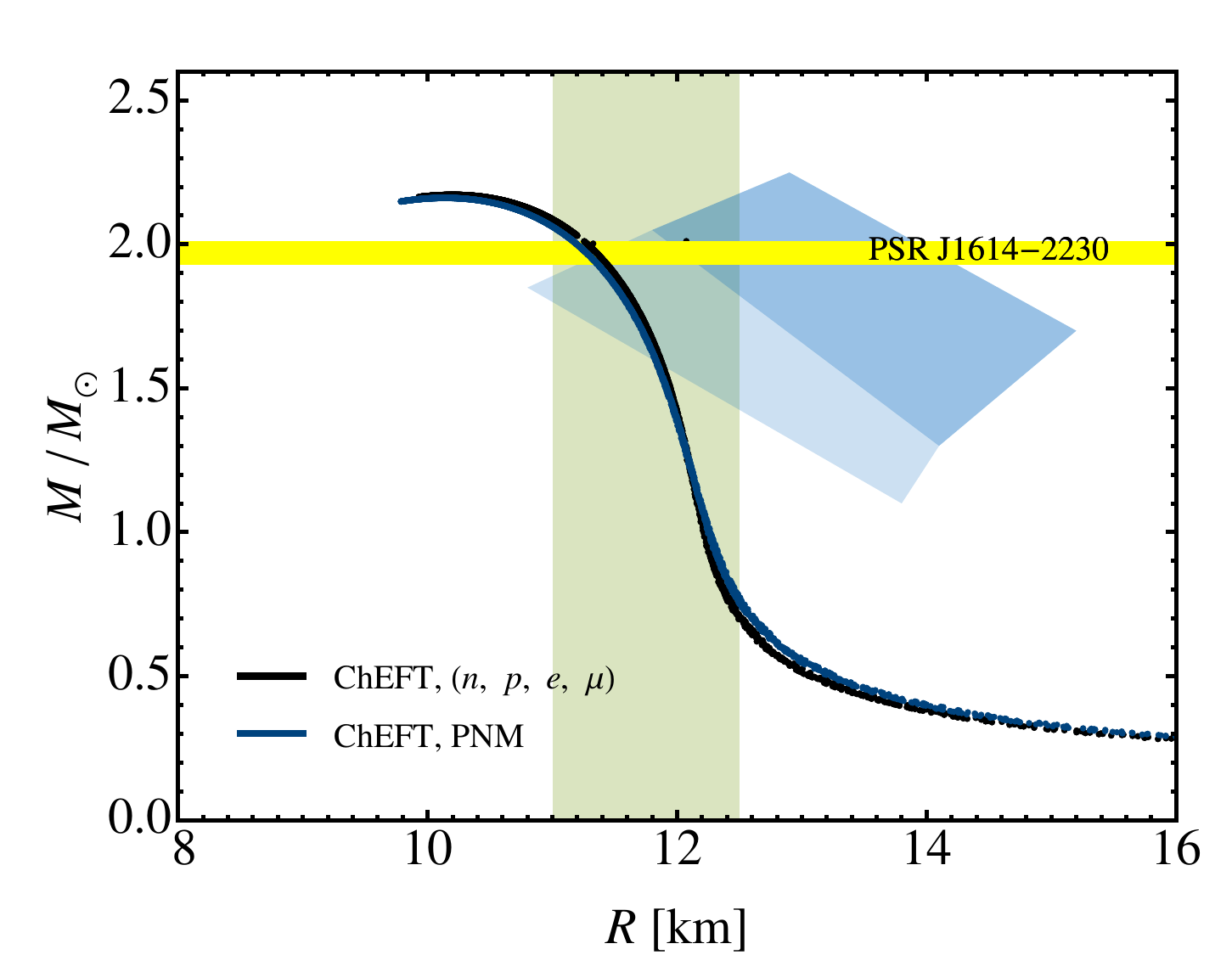}
	\end{minipage}
\vspace{0.4cm} 

\sidecaption
\caption{Left: EoS by using the ChEFT ansatz described in the text with inclusion of beta-equilibrium and charge-neutrality conditions (black line) and for pure neutron matter (blue line). The purple line  is the EoS for beta-stable nuclear matter obtained from a variational many-body calculation by Akmal, Pandharipande, Ravenhall \cite{Akmal:1998cf}. Right: $M=M(R)$ of the corresponding scenarios.  Green and blue bands: constraints on the EoS, the masses and the radii as given in  \cite{Steiner:2010fz,Steiner:2012xt} and \cite{Truemper2013}, respectively.}
\label{fig-eoschpt} 
\end{figure}

{\bf  Matter composed of nucleons and mesons}\label{sec-41}. The EoS for asymmetric nuclear matter is constructed by combining symmetric matter and pure neutron matter, each calculated using in-medium ChEFT, in a quadratic expansion of the energy per particle in the asymmetry parameter: $\bar E_{\rm asym}=\bar E_{\rm sym}(\varrho)\left(1-\delta^2\right)+\bar E_{\rm PNM}(\varrho)\,\delta^2$, with $\delta=\frac{\varrho_n-\varrho_p}{\varrho}$.	The resulting EoS (pressure as function of energy density), including charge neutrality and beta equilibrium, is shown in the left panel of Fig.~\ref{fig-eoschpt}.
The mass-radius relation for neutron stars based on this EoS is displayed in the right panel of Fig.~\ref{fig-eoschpt}. It turns out that the ChEFT EoS with its important three-body interactions and explicit treatment of Pauli effects in two-pion exchange processes, produces a sufficiently stiff EoS in order to support two-solar-mass neutron stars, even in the absence
of more exotic material in the center of the star. The central density of a typical two-solar-mass neutron star does not exceed about $5\,\varrho_0$. While this raises questions about the convergence of the ChEFT expansion, one should note that the typical nucleon Fermi momenta $k_F$ under these conditions are still not larger than about $4m_\pi$ and much smaller than the spontaneous chiral symmetry breaking scale, $4\pi f_\pi\sim 1$ GeV.

\begin{figure}
\begin{minipage}{.47\textwidth}\vspace{.65cm}
		\includegraphics[width=\textwidth]{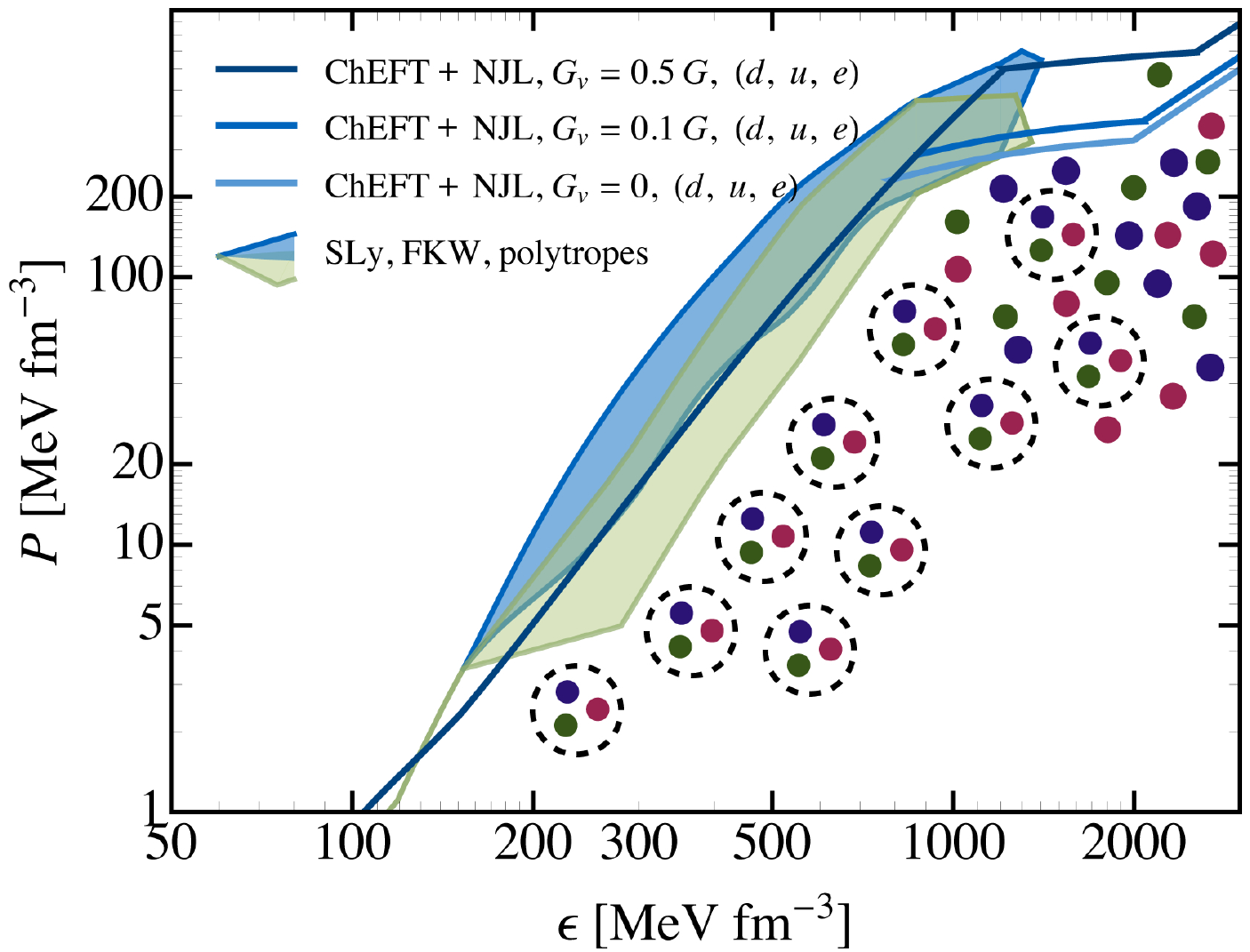}
	\end{minipage}
	\hspace{0.5cm}
\begin{minipage}{.465\textwidth}
		\vspace{0.55cm}
		\includegraphics[width=\textwidth]{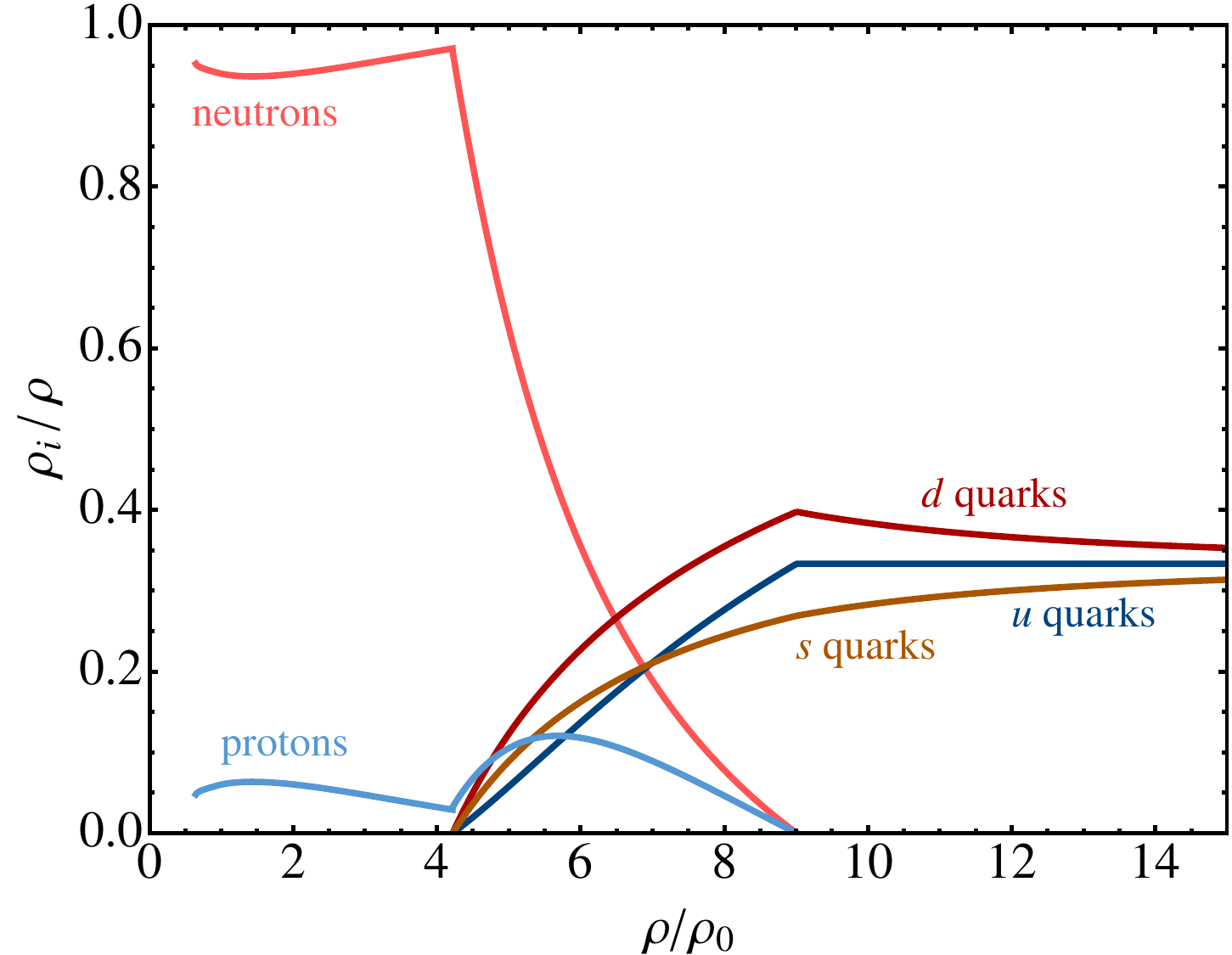}
	\end{minipage}

\vspace{0.4cm} 
\sidecaption
\caption{Left: EoS for different vector coupling strengths. At $G_v=0$ the coexistence region is $4\varrho_0\le \varrho\le 9\varrho_0$. 
Right:  particle ratios, $\rho_i/\rho$, as a function of the baryon density $\rho/\rho_0$. } \label{fig-njlgibbs}
\end{figure}

{\bf Hybrid stars}\label{sec-42}. The EoS for possible hybrid stars including a quark-matter core is studied first adopting  the picture of a quark-hadron crossover \cite{Schafer:1998ef,Baym:2008me,Maeda:2009ev}, assuming that hadrons at high densities begin to overlap such that a distinction between hadrons and quarks becomes meaningless. This scenario is modeled \cite{Masuda:2012ed} by interpolating smoothly between the ChEFT and the NJL EoS. In this case it turns out that strong vector repulsion in the NJL interaction between quarks is needed in order to meet the stiffness requirement for the EoS, as already pointed out in Ref.~\cite{Masuda:2012ed}.
As a second possible scenario we consider a first-order transition between baryonic matter and NJL-type quark matter. In addition to chemical equilibrium and charge neutrality, the Gibbs conditions \cite{Glendenning:1992vb} for thermal and mechanical equilibrium, $\label{gibbs} P_{\rm H}(\mu_n,\mu_e)=P_{\rm QM}(\mu_n,\mu_e)$\,, are imposed, where $P_{\rm H}$ and $P_{\rm QM}$ denote the pressure in the hadronic and in the quark-matter phases, respectively. Using this ansatz there is no ambiguity about where the coexistence region arises. Figure~\ref{fig-njlgibbs} shows the resulting EoS's for different NJL vector-coupling strengths $G_v$ (left panel). The corresponding mass-radius relation looks very similar to the one shown in Fig.~\ref{fig-eoschpt} (right). A quark-nuclear coexistence region now appears at baryon densities above $4\,\varrho_0$. This is also where strange quarks begin to appear in this picture. The particle distributions (in this case for $G_v=0$) are shown in Fig.~\ref{fig-njlgibbs} (right panel). Choosing $G_v > 0$ moves the onset of quark matter to even higher densities.

\section{Discussion and conclusions}\label{sec-5}

We have investigated the equation of state of baryonic matter using constraints from recent neutron-star observations. It is found that an EoS derived from ChEFT with baryonic and mesonic (dominantly pionic) degrees of freedom and including three-body forces is compatible with observation, although questions about higher order terms beyond three-body are still open. On the
other hand, the central density in a two-solar-mass neutron star does not exceed about $5\,\varrho_0$ with this EoS, so that the average inter-nucleon distance is still larger than 1 fm and there is no significant overlap of the valence quark cores in the nucleons. Concerning an inner core composed of quark matter, a first-order transition would lead to quark-hadron coexistence at baryon densities exceeding $\varrho\gtrsim 4\,\varrho_0$, rendering such a scenario not very significant even for the deep center of the star. Strange quarks appear at such densities and leave not much room for hyperons which would soften the EoS unless strong short-range repulsions come into play again. Our present analysis strongly supports, however, a very stiff equation of state without the need for substantial exotic-matter admixtures, in order to reproduce
the empirical observations for neutron stars.


\begin{thebibliography}{}


\bibitem{Demorest:2010bx}
P.~Demorest, T.~Pennucci, S.~Ransom, M.~Roberts, J.~Hessels, Nature
  \textbf{467}, 1081 (2010)

\bibitem{Antoniadis:2013pzd}
J.~Antoniadis, P.C. Freire, N.~Wex, T.M. Tauris, R.S. Lynch et~al., Science
  \textbf{340}, 6131 (2013)

\bibitem{Lattimer:2010uk}
J.M. Lattimer, M.~Prakash (2010), \texttt{1012.3208}

\bibitem{Steiner:2010fz}
A.W. Steiner, J.M. Lattimer, E.F. Brown, Astrophys.J. \textbf{722}, 33 (2010)

\bibitem{Steiner:2012xt}
A.W. Steiner, J.M. Lattimer, E.F. Brown, Astrophys.J.Lett. \textbf{765}, L5
  (2013)

\bibitem{Lattimer:2013hma}
J.M. Lattimer, A.W. Steiner (2013), \texttt{1305.3242}

\bibitem{Guillot:2013wu}
S.~Guillot, M.~Servillat, N.A. Webb, R.E. Rutledge (2013), \texttt{1302.0023}

\bibitem{Truemper2013}
J.E. Tr{\"u}mper, Prog.Part.Nucl.Phys. \textbf{66}, 674 (2011)

\bibitem{Hebeler:2010jx}
K.~Hebeler, J.~Lattimer, C.~Pethick, A.~Schwenk, Phys.Rev.Lett. \textbf{105},
  161102 (2010)

\bibitem{Hebeler:2013nza}
K.~Hebeler, J.~Lattimer, C.~Pethick, A.~Schwenk (2013), \texttt{1303.4662}

\bibitem{Haensel:2004nu}
P.~Haensel, A.Y. Potekhin, Astron.Astrophys. \textbf{428}, 191 (2004)

\bibitem{Douchin:2001sv}
F.~Douchin, P.~Haensel (2001), \texttt{astro-ph/0111092}

\bibitem{Fiorilla:2011sr}
S.~Fiorilla, N.~Kaiser, W.~Weise, Nucl.Phys. \textbf{A880}, 65 (2012)

\bibitem{Kaiser:2001jx}
N.~Kaiser, S.~Fritsch, W.~Weise, Nucl.Phys. \textbf{A697}, 255 (2002); Nucl.Phys. \textbf{A750}, 259 (2005)

\bibitem{Kaiser:2011cg}
N.~Kaiser, Nucl.Phys. \textbf{A860}, 41 (2011)


\bibitem{Vogl:1991qt}
U.~Vogl, W.~Weise, Prog. Part. Nucl. Phys. \textbf{27}, 195 (1991)

\bibitem{Klimt:1989pm}
S.~Klimt, M.~Lutz, U.~Vogl, W.~Weise, Nucl. Phys. \textbf{A516}, 429 (1990); 
S.~Klimt, M.~Lutz, W.~Weise, Phys. Lett. \textbf{B249}, 386 (1990)

\bibitem{Akmal:1998cf}
A.~Akmal, V.~Pandharipande, D.~Ravenhall, Phys.Rev. \textbf{C58}, 1804 (1998)

\bibitem{Schafer:1998ef}
T.~Sch{\"a}fer, F.~Wilczek, Phys.Rev.Lett. \textbf{82}, 3956 (1999)

\bibitem{Baym:2008me}
G.~Baym, T.~Hatsuda, M.~Tachibana, N.~Yamamoto, J.Phys. \textbf{G35}, 104021
  (2008)

\bibitem{Maeda:2009ev}
K.~Maeda, G.~Baym, T.~Hatsuda, Phys.Rev.Lett. \textbf{103}, 085301 (2009)

\bibitem{Masuda:2012ed}
K.~Masuda, T.~Hatsuda, T.~Takatsuka (2012), \texttt{1212.6803}

\bibitem{Glendenning:1992vb}
N.K. Glendenning, Phys.Rev. \textbf{D46}, 1274 (1992)




%
%
%
\end{thebibliography}
\end{document}